\documentclass[aip,jap,groupedaddress,12pt]{revtex4-1}

\usepackage{graphicx}

\usepackage{hyperref}
\hypersetup{colorlinks=false}

\begin{document}

\title{Self-injection-locked magnetron as an active ring resonator side coupled to a waveguide with a delayed feedback loop}

\author{Y.P. Bliokh}
\affiliation{Department of Physics, Technion, 32000 Haifa, Israel}
\author{Ya.E. Krasik}
\affiliation{Department of Physics, Technion, 32000 Haifa, Israel}
\author{J. Felsteiner}
\affiliation{Department of Physics, Technion, 32000 Haifa, Israel}

\begin{abstract}
The theoretical analysis and numerical simulations of the magnetron operation with a feedback loop were performed assuming that the delay of the electromagnetic wave propagating in the loop is constant whereas the phase of the complex feedback reflection coefficient is varied. Results of simulations showed that by a proper adjustment of values of the time delay and phase of reflection coefficient that determines phase matching between the waves in the resonator and feedback loop, one can increase the magnetron's output power significantly without any other additional measures.

\end{abstract}

\pacs{}

\maketitle

\section{Introduction}

Magnetron oscillators are widely used sources of microwave radiation operating over a wide range of powers and frequencies and having various applications \cite{Magnetron}. A typical feature of the magnetron operation is a relatively broad and noisy spectrum \cite{Noise}. However, for many applications (radar, communication systems, charged particles accelerators, etc.), it is very necessary both to narrow the frequency spectrum and to reduce the noise. An improvements in the microwave spectrum can be achieved either by using a special magnetron design, which makes starting conditions more preferable for the operating mode, or by feeding external monochromatic electromagnetic radiation into the magnetron cavity. In the latter case, this controlling radiation stabilizes the magnetron operation similarly to other nonlinear oscillators  \cite{Adler}. In fact, the injected electromagnetic radiation can be produced by either an external source [injection-locked magnetron (ILM)\cite{Weglein, Treado, Tahir, Dexter}]  or by the same magnetron [self-inject-locked magnetron (SILM)\cite{Choi}]. In the latter case, a part of the magnetron output power is injected back into the magnetron cavity via a feedback loop. It is worth noting that this feedback loop can appear as a result of the magnetron radiation reflection from the load. It had been shown \cite{Weglein, Treado, Tahir, Dexter,Choi} that ILMs and SILMs are able to fix the operating frequency and phase, and essentially decrease the noise of the radiation. 

Recent investigations \cite{Krasik-1, Krasik-2} show that the efficiency of the relativistic S-band magnetron can be significantly (up to 40\%) increased when part of the radiated electromagnetic power is reflected into the magnetron cavity. This effect, to our knowledge, has not yet been discussed. In this paper we suggest an explanation for this effect using a model of the magnetron which considers it as an active nonlinear ring resonator. This model can be applied also to coupled ring resonator-fiber systems which are of interest for optoelectronics and communication (see, e.g., \cite{Little, Yariv_2}).

\section{Geometry of the problem \label{Geometry}}

\begin{figure}[htb]
\centering \scalebox{0.35}{\includegraphics{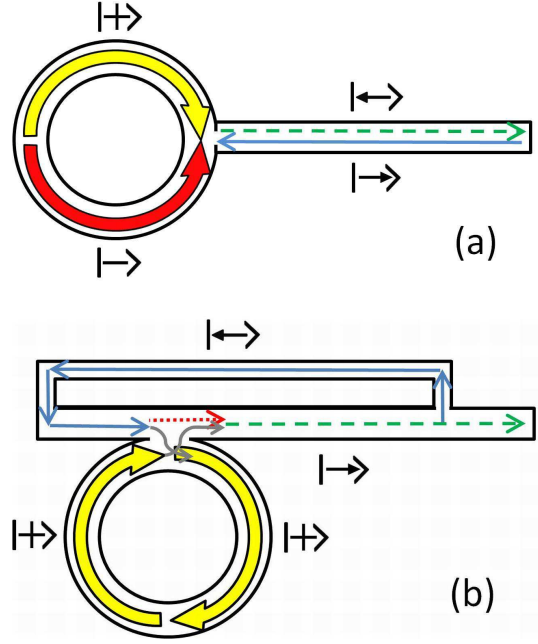}}
\caption{Magnetron (ring resonator) and waveguide. (a) -- Clockwise $|+\rangle$ (yellow bold arrow) and anti-clockwise $|-\rangle$) (red bold arrow) propagating eigenmodes. Direct (dashed line) and reflected (solid line) waves in the waveguide. (b) -- Ring resonator side-coupled to a waveguide with a feedback loop.} \label{Fig1}
\end{figure}

In the model, a magnetron is considered as a ring resonator (see Fig.~\ref{Fig1})with double degenerated eigenmodes, clockwise $|+\rangle$ and anti-clockwise $|-\rangle$, propagating waves having equal frequencies $\omega$.  Only one wave  (mode $|+\rangle$, for definiteness) can be excited due to the resonant interaction with the rotating electron flow when the electron azimuthal drift velocity is close to the wave phase velocity. The energy stored in the $|+\rangle$ wave (hereafter active mode) leaks partially from the resonator to the waveguide (see Fig.~\ref{Fig1}a).

The output end of the waveguide is connected to a certain payload, for instance, an antenna. It is assumed that the matching between the waveguide and the load is not ideal, i.e., part of the wave energy is reflected and enters back into the magnetron cavity. This reflected wave excites two contra-propagating waves, $|+\rangle$ and $|-\rangle$, in the magnetron cavity. The active, $|+\rangle$, wave is amplified by the electron flow, whereas the passive wave, $|-\rangle$, propagates without amplification. Thus, one can neglect the passive wave and consider the waves' paths as shown in Fig.~\ref{Fig1}b. By convention, the paths of rightward and leftward propagating waveguide modes ($|\rightarrow\rangle$ and $|\leftarrow\rangle$ modes) can be considered as separated paths. Namely, the wave which is reflected from the waveguide output end comes back to the resonator along a \textit{feedback loop} (see Fig.~\ref{Fig1}b). In this scheme, the ring resonator is \textit{side-coupled} to the waveguide. The coupling between the resonator and waveguide modes is similar to a \textit{phase matched} coupling. This means that only the co-propagating waves are coupled: modes $|\rightarrow\rangle$ and $|+\rangle$ (modes $|\leftarrow\rangle$ and $|-\rangle$) are coupled, whereas modes $|\rightarrow\rangle$ and $|-\rangle$ (modes $|\leftarrow\rangle$ and $|+\rangle$) are not coupled.  This configuration, i.e., side-coupled waveguide and ring (or disk) resonator, has widely been explored in photonics (see, e.g., \cite{Little, Yariv_2}). The general properties of this system were reported in Ref. \cite{Yariv_1} and the results of this work will be used in the next section.

\section{Small signal model \label{Linear}}

\begin{figure}[htb]
\centering \scalebox{0.35}{\includegraphics{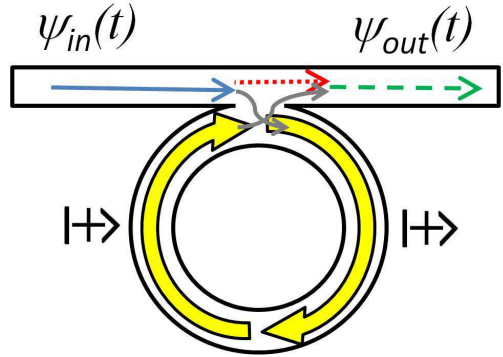}}
\caption{Auxiliary problem. Phase matched coupling between waveguide and waveguide modes. Notation are the same as in Fig.~\ref{Fig1}} \label{Fig2}
\end{figure}

Let us consider a supplementary problem, namely wave propagation in the waveguide coupled with a ring resonator (see Fig.~\ref{Fig2}). 
Let $\psi_{in,\omega}$  be the complex amplitude 
of the monochromatic rightward propagating waveguide mode $\psi_{in}(t)=\psi_{in,\omega}e^{-i\omega t}$ on the left side with respect to the waveguide-resonator coupling cross-section. The complex amplitude $\psi_{out,\omega}$  of the transmitted wave $\psi_{out}(t)=\psi_{out,\omega}e^{-i\omega t}$  on the right side with respect to this cross-section and amplitude  are related as: 
\begin{equation}\label{eq1}
\psi_{out,\omega}=t(\omega)\psi_{in,\omega},
\end{equation} 
where the transmission coefficient $t(\omega)$ is defined as \cite{Yariv_1}:
\begin{equation}\label{eq2}
t(\omega)=1-i{1\over \omega-\omega_r+i\Gamma}{L\over
v_g}\left|V_{r,w}\right|^2.
\end{equation}
Here $\omega_r$ is the eigenfrequency of the ring resonator, $\Gamma=\Gamma_d+\Gamma_c$, the coefficient $\Gamma_d$ is related to the resonator dissipative losses, the coefficient $\Gamma_c=\left|V_{r,w}\right|^2 L/2v_g$ is related to the energy leakage from the resonator because of coupling with a waveguide characterized by the coupling coefficient $V_{r,w}$, $L$ is the resonator perimeter, $v_g$ is the group velocity of the resonator eigenmode. Eq.~(\ref{eq2}) is valid when the coupling is weak: 
\begin{equation}\label{eq3}
\Gamma_c/\omega_r\ll 1.
\end{equation}

Parameters $\Gamma$, $\Gamma_d$, and  $\Gamma_c$ can be written in terms of the resonator $Q$-factors: $\Gamma=\omega_r/2Q_{tot}$,
$\Gamma_d=\omega_r/2Q_{diss}$, and $\Gamma_c=\omega_r/2Q_{leak}$. Here $Q_{tot}$ is the total $Q$-factor, and $Q_{diss}$ and $Q_{leak}$ are the $Q$-factors responsible for the energy dissipation and leakage, respectively. A \textit{negative} dissipative $Q$-factor, $Q_{diss}<0$, corresponds to an \textit{active} resonator, which amplifies an incoming wave.

The output wave amplitude $\psi_{out,\omega}$ is a superposition of two amplitudes, namely, the amplitude of the wave which crosses the coupling region directly through the waveguide, and the amplitude of the wave that enters the waveguide from the resonator. The two terms in the right side of  Eq.~(\ref{eq2}) describe these two paths of the wave transmission. The first term, unity, describes the wave's direct transmission through the waveguide \cite{Note_1}; the second term describes the wave's transmission through the resonator. Thus, the output wave amplitude $\psi_{out,\omega}$ is the sum of two amplitudes: 
\begin{equation}\label{eq4}
\psi_{out,\omega}=t(\omega)\psi_{in,\omega}=\psi_{direct,\omega}+\psi_{res,\omega},
\end{equation}
where $\psi_{direct,\omega}\simeq \psi_{in,\omega}$ is the amplitude of the wave transmitted directly through the waveguide, and $\psi_{res,\omega}$ is the amplitude of the wave that transmitted through the resonator:
\begin{equation}\label{eq5}
\psi_{res,\omega}=[t(\omega)-1]\psi_{in,\omega}=
-{iQ_{leak}^{-1}\over\omega/\omega_r-1+i\left(Q_{diss}^{-1}+Q_{leak}^{-1}\right)/2}\psi_{in,\omega}.
\end{equation}

Eq.~(\ref{eq5}) is  the Fourier transformation of the differential equation:
\begin{equation}\label{eq6}
{d\psi_{res}(t)\over dt}+i\omega_r\psi_{res}(t)+{\omega_r\over
2}\left({1\over Q_{leak}}+{1\over
Q_{diss}}\right)\psi_{res}(t)=-{\omega_r\over Q_{leak}}\psi_{in}(t).
\end{equation}
Thus, one can write the following equation for input and output waves having arbitrary shapes: 
\begin{equation}\label{eq7}
\psi_{out}(t)=\psi_{in}(t)+\psi_{res}(t),
\end{equation}
where $\psi_{res}(t)$ is the solution of Eq.~(\ref{eq6}).
Here, let us remind that the wave $\psi_{res}(t)$ is the wave that leaks from the resonator into the waveguide. The wave field in the resonator can be significantly larger than the wave field $\psi_{res}(t)$  in the waveguide [weak coupling between the resonator and waveguide, see Eq.~(\ref{eq3}]. However, these wave fields are linearly proportional to each other and the field $\psi_{res}(t)$  can be considered as a normalized field in the resonator.

A supplementary problem that is being considered helps one to describe the system with the presence of a feedback loop (see Fig.~\ref{Fig1}b). In a general case, the wave propagation along the feedback loop is described by a functional equation: 
\begin{equation}\label{eq8}
\psi_{in}(t)=\hat{F}\left\{\psi_{out}(t)\right\},
\end{equation}
where $\hat{F}$ is a certain functional. In the simplest case of a non-dispersive feedback loop this relation has the form:
\begin{equation}\label{eq9}
\psi_{in}(t)=q\psi_{out}(t-T),
\end{equation}
where $T$ is the time of the wave propagation along the loop (delay time), and $q$ is the complex feedback coefficient, $|q|<1$. Below we will consider this simple model only.

Eqs.~(\ref{eq6}), (\ref{eq7}), and (\ref{eq9}) [(\ref{eq8}) in the general case] being supplemented by initial conditions, are a closed system that describes the small signal theory of a magnetron with a delayed feedback. Asymptotic solutions of these equations are rather simple. There is an exponentially either growing or decaying output wave amplitude, depending on whether the system's self-excitation threshold is exceeded or not. In order to describe the wave amplitude saturation, non-linear terms should be incorporated in these equations.

\section{Nonlinear model \label{Nonlinear}} 

Without the feedback loop, i.e.,  when $q=0$ ($\psi_{in}=0$), Eq.~(\ref{eq6}) describes a linear resonator with the self-excitation threshold defined as:
\begin{equation}\label{eq10}
Q_{diss}+Q_{leak}=0.
\end{equation}

When $Q_{diss}^{-1}+Q_{leak}^{-1}<0$, the field amplitude $\psi_{res}(t)$ growths exponentially. The increase in the amplitude can be limited by the replacement of the \textit{linear} dissipation term $Q_{diss}^{-1}<0$ in Eq.~(\ref{eq6}) by the \textit{nonlinear} one:  $Q_{diss}^{-1}\rightarrow\tilde{Q}_{diss}^{-1}=Q_{diss}^{-1}\left(1-|\psi_{res}(t)|^2/\gamma_d^2\right)$, where $\gamma_d$ is the parameter that is related to the saturation amplitude. With nonlinear $Q_{diss}^{-1}$ Eq.~(\ref{eq6}) takes the form:
\begin{equation}\label{eq11}
{d\psi_{res}(t)\over dt}+i\omega_r\psi_{res}(t)+{\omega_r\over
2}\left[{1\over Q_{leak}}+{1\over
Q_{diss}}\left(1-|\psi_{res}(t)|^2/\gamma_d^2\right)\right]\psi_{res}(t)=0.
\end{equation}
This is the Van der Pol equation which is widely used as a model of a nonlinear auto-oscillating dynamical system \cite{Textbook}. Another effect which, in general, should be taken into account, is a nonlinear shift of the resonator eigenfrequency. This effect can be incorporated for in the model by the replacement in Eq.~(\ref{eq6}) $\omega_r\rightarrow\omega_r\left(1-|\psi_{res}(t)|^2/\gamma_f\right)$, where $\gamma_f$ is the parameter that is related to the frequency shift. The latter results in describing the Duffing oscillator. Both Van der Pol and Duffing models are typical examples of nonlinear two-dimensional oscillating systems \cite{Textbook}. Duffng's model describes non-isochronal oscillations, and Van der Pol's model describes a limit cycle when the equilibrium point loses its stability. It is reasonable to suppose that, due to their generality, these models can be used to describe the properties of the system under consideration. To summarize, let us present below the equations which will be analyzed: 
\begin{eqnarray}
{d\psi_{res}(t)\over
dt}+i\omega_r\left(1-|\psi_{res}(t)|^2/\gamma_f\right)\psi_{res}(t)+\nonumber\\
{\omega_r\over 2}\left[{1\over Q_{leak}}+{1\over
Q_{diss}}\left(1-|\psi_{res}(t)|^2/\gamma_d^2\right)\right]\psi_{res}(t)=
-{\omega_r\over Q_{leak}}\psi_{in}(t),\label{eq12}\\
\psi_{in}(t)=q\psi_{out}(t-T),\label{eq13}\\
\psi_{out}(t)=\psi_{in}(t)+\psi_{res}(t).\label{eq14}
\end{eqnarray}

\section{Analysis \label{Analysis}}

One can rewrite Eqs.~(\ref{eq12})-(\ref{eq14}) in the dimensionless form using dimensionless time $\tau=\omega_rt$ and normalized amplitudes $A_i(\tau)=\psi_i(\tau) e^{i\tau}\gamma_d^{-1}\sqrt{Q_{leak}/\left(Q_{leak}-|Q_{diss}|\right)}$,:
\begin{eqnarray}
{dA_{res}(\tau)\over d\tau}-i\kappa |Q_{diss}|{\gamma_d^2\over
\gamma_f}|A_{res}(\tau)|^2A_{res}(\tau)-\nonumber\\ 
{1\over 2}\kappa\left[1-
|A_{res}(\tau)|^2\right]A_{res}(\tau)=
-{1\over Q_{leak}}A_{in}(\tau),\label{eq15}\\
A_{in}(\tau)=qe^{-i\theta}A_{out}(\tau-\theta),\label{eq16}\\
A_{out}(\tau)=A_{in}(\tau)+A_{res}(\tau),\label{eq17}
\end{eqnarray}
Here $\theta=\omega_rT$ is the dimensionless delay time and  $\kappa=\left(Q_{leak}-|Q_{diss}|\right)/ Q_{leak}|Q_{diss}|$. In Eq.~(\ref{eq15}) it was taken into account that $Q_{diss}<0$.

Eqs.~(\ref{eq15})-(\ref{eq17}) have monochromatic solutions with constant amplitudes. Setting $A_i(\tau)=B_ie^{-i\nu\tau}$, where $\nu=(\omega-\omega_r)/\omega_r$ is the normalized frequency shift, we arrive at the following set of algebraic equations:
\begin{eqnarray}
-i\nu B_{res}-i\kappa |Q_{diss}|{\gamma_d^2\over
\gamma_f}|B_{res}|^2B_{res}- {1\over 2}\kappa\left(1-
|B_{res}|^2\right)B_{res}=
-{1\over Q_{leak}}B_{in},\label{eq18}\\
B_{in}=qe^{-i\theta+i\nu\theta}B_{out},\label{eq19}\\
B_{out}=B_{in}+B_{res}.\label{eq20}
\end{eqnarray}
Excluding $B_{in}$, one can obtain:
\begin{equation}\label{eq21}
i\left(\nu+\kappa |Q_{diss}|{\gamma_d^2\over
\gamma_f}|B_{res}|^2\right)+{1\over 2}\kappa\left(1-
|B_{res}|^2\right)={1\over Q_{leak}}{qe^{-i\theta+i\nu\theta}\over
1-qe^{-i\theta+i\nu\theta}}.
\end{equation}
Eq.~(\ref{eq21}) determines both frequency shift $\nu$ and normalized power in the resonator $P=|B_{res}|^2$ ($P=1$ without the feedback, $q=0$). In the simple case when the nonlinear dispersion absent $(\gamma_f\rightarrow\infty)$ one can rewrite the complex equation (\ref{eq21}) as:
\begin{eqnarray}
\nu=-{|q|\over Q_{leak}}{\sin(\theta-\nu\theta-\alpha)\over
1+|q|^2-2|q|\cos(\theta-\nu\theta-\alpha)},\label{eq22}\\
P=|B_{res}|^2=1-{2|q|\over\kappa
Q_{leak}}{\cos(\theta-\nu\theta-\alpha)-|q|\over
1+|q|^2-2|q|\cos(\theta-\nu\theta-\alpha)},\label{eq23}
\end{eqnarray}
where $\alpha=\arg q$ the phase of the feedback coefficient.

As it follows from Eq.~(\ref{eq23}), the normalized power can be varied in the range:
\begin{equation}\label{eq24}
1-{2|q|\over 1-|q|}{1\over\kappa Q_{leak}}\leq P\leq 1+{2|q|\over
1+|q|}{1\over \kappa Q_{leak}}.
\end{equation}  
When the left boundary in (\ref{eq24}) is negative, the minimal possible power is equal to zero.

Near the self-excitation threshold, the parameter $\kappa Q_{leak}=\left(Q_{leak}-|Q_{diss}|\right)/|Q_{diss}|$ is small. In the latter case the feedback can noticeably increases the power $P$ in the resonator.
The inequalities (\ref{eq24}) determine \textit{potentially-possible} boundaries of the power variation. Upper and lower boundaries correspond to such values of the frequency shift $\nu$  for which $\cos(\theta-\nu\theta-\alpha)=\mp 1$, respectively. However, at designated values $\theta$ and $\alpha$, the frequency $\nu$ is determined by the solution of the \textit{dispersion equation} (\ref{eq22}). The proximity of the power value to the upper possible value depends on whether the solutions of the dispersion equation contain eigenfrequencies $\nu$ for which $\cos(\theta-\nu\theta-\alpha)$ is rather close to $-1$.

It is convenient to present the dispersion equation (\ref{eq24}) as:
\begin{equation}\label{eq25}
\varphi=\theta-\alpha+{\theta|q|\over Q_{leak}}{\sin\varphi\over
\left(1+|q|^2-2|q|\cos\varphi\right)},
\end{equation}
where $\varphi=\theta-\alpha-\theta\nu$. If the feedback coefficient is small, $|q|^2\ll 1$, and
\begin{equation}\label{eq26}
\theta<Q_{leak}/|q|,
\end{equation}
Eq.~(\ref{eq25}) has one solution $\varphi(\alpha)$ (assuming delay time $\theta={\rm const}$). The inequality (\ref{eq26}) ) will be considered as a \textit{short} delay time condition. In the case of a reflection coefficient phase $\alpha$ variation, the value $\cos[\varphi(\alpha)]$ varies in a full range from $-1$ to $+1$.
Accordingly, the power $P$ varies in a full range defined by Eq.~(\ref{eq23}). In the opposite case of a \textit{long} delay time , when
\begin{equation}\label{eq27}
\theta\gg Q_{leak}/|q|,
\end{equation}
the dispersion equation (\ref{eq25}) has many roots, and the eigenfrequency $\nu$ for which $\cos[\theta-\nu(\alpha)\theta-\alpha]$ is close to $-1$ is always present. This means that Eqs.~(\ref{eq22}) and (\ref{eq23}) always have  solution for which the power $P$ value is close to its upper limit defined by Eq.~(\ref{eq24}).

The analysis presented above shows that a variation in the reflection coefficient phase can produce 
a rather strong power variation, if the inequality (\ref{eq26}) is satisfied.  In the opposite case, when the inequality (\ref{eq27}) is satisfied, the power is close to its maximal possible value, which can noticeably exceed the power in the resonator without a feedback loop.

\section{Numerical simulations \label{Numerical}}   

\begin{figure}[h]
\centering \scalebox{1.0}{\includegraphics{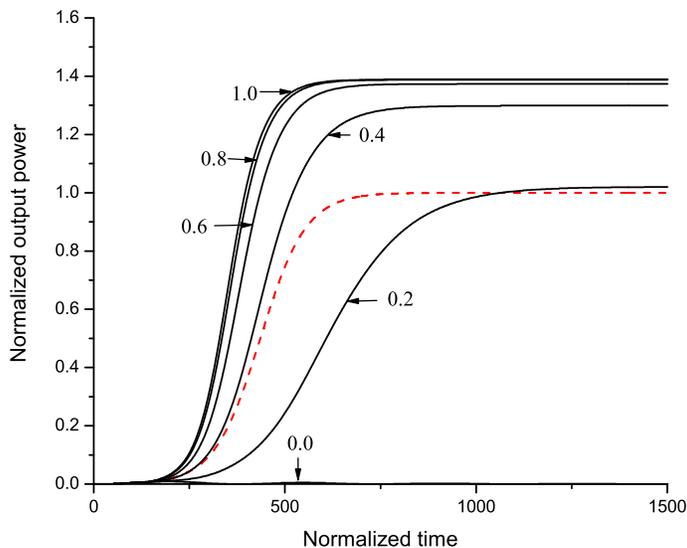}}
\caption{Dependence of the normalized power on time. ``Short'' delay time $\theta<Q_{leak}/|q|$. Red dashed line corresponds to $q=0$. The curves are marked by corresponding $\alpha/\pi$ values. $|Q_{diss}|= 15$, $Q_{leak}= 20$, $q= 0.2$, $\theta= 70$.} \label{Fig3}
\end{figure}

\begin{figure}[h]
\centering \scalebox{1.0}{\includegraphics{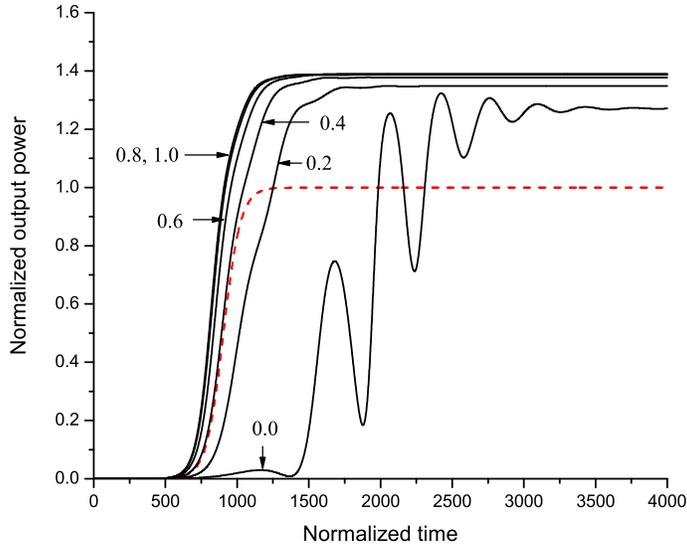}} \caption{The same as in Fig.~\ref{Fig3}. ``Long'' feedback loop, $\theta\gg Q_{leak}/|q|$. $|Q_{diss}|= 15$, $Q_{leak}= 20$, $q= 0.2$, $\theta= 200$.} \label{Fig4}
\end{figure}

Examples of numerical solutions of Eqs.~(\ref{eq15})-(\ref{eq17}) are presented in Figs.~\ref{Fig3} and \ref{Fig4}. 

Fig.~\ref{Fig3} demonstrates how the power saturation level depends on the phase $\alpha$ when the inequality (\ref{eq26}) is satisfied. Parameters used for numerical simulation were chosen close to their estimations for the experimental setup described in Refs.~\cite{Krasik-1,Krasik-2}: $|Q_{diss}|= 15$, $Q_{leak}= 20$, $q= 0.2$, $\theta= 70$. The results of simulations showed that the microwave generation in the magnetron can be either totally suppressed ($\alpha=0$) or enhanced by $\sim 40$\% ($\alpha\simeq\pi$) as compared with the magnetron operation without feedback ($q=0$).

Fig.~\ref{Fig4} is related to the ``long'' feedback loop when the inequality (\ref{eq27}) is satisfied. In agreement with the analysis presented above, the power saturation level is almost independent of the phase $\alpha$ and is close to its maximal value for the system with a ``short'' feedback loop (short delay time).

\section{Discussion}

The theoretical analysis and numerical simulations of the magnetron operation with a feedback loop were performed assuming that the time $\theta$ of the wave propagating in the feedback loop is constant whereas the phase $\alpha$ of the reflection coefficient $q$ is varied. These assumptions were used only to simplify the solution. In general, the solution of the dispersion equation (\ref{eq25}) depends on both phase $\alpha$ and delay time $\theta$. Variation in the value of $\theta$ affects the solution in the same manner as the phase $\alpha$ variation. This similarity becomes important particularly when $\theta\gg\pi$  and the condition (\ref{eq26}) is satisfied. Thus, the length of the waveguide that connects the magnetron with a payload (an antenna, for instance) can affect the magnetron characteristics essentially. Results of simulations showed that by a proper adjustment of values of $\theta$ and $\alpha$ that determines phase matching between the waves in the resonator and feedback loop, one can increase the magnetron's output power significantly without any other additional measures.


\begin{thebibliography}{99}
\bibitem{Note_1} The fraction $\sim Q_{leak}^{-1}$ of the incoming wave energy passes through the resonator, therefore corresponding transmission coefficient, $t_w$ is not equal to unity. This fraction is small by virtue of condition (\ref{eq3}) and can be neglected because both terms in Eq.~(\ref{eq2}) are the same order of magnitude, of the order of unity.
\end{thebibliography}
\end{document}